\documentclass[12pt]{article}

\usepackage{epsfig}

\begin{document}

\title{\bf \large MODELING COLLECTIVE DISLOCATION DYNAMICS IN ICE
SINGLE CRYSTALS}

\author{\normalsize M.-CARMEN MIGUEL$^{*}$, A. VESPIGNANI$^{*}$, and
S. ZAPPERI$^{**}$ \\ \normalsize $^*$The Abdus Salam International
Centre for Theoretical Physics
\\ \normalsize P.O. Box 586, 34100 Trieste, Italy\\
\normalsize $^{**}$PMMH-ESPCI,10, rue Vauquelin, 75231 Paris Cedex 05,
France}

\date{}

\maketitle

\thispagestyle{empty}

\section*{\bf \normalsize ABSTRACT}

We propose a model to study the plasticity of ice single crystals by
numerical simulations. The model includes the long-range character of
the interaction among dislocations, as well as the possibility of
mutual annihilation of these line defects characterized by its Burgers
vector. A multiplication mechanism representing the activation of
Frank-Read sources due to dislocation pinning is also introduced in
the model.

With our approach we are able to probe the dislocation patterns, which
result from the dislocation dynamics. Furthermore, our results exhibit
features characteristic of driven dynamic critical phenomena such as
scaling behavior, and avalanche dynamics. Some of these results
account for the experimental findings reported for ice single crystals
under creep deformation, like the power-law distributions of the
acoustic emission intensity observed sistematically in experiments.

\section*{\bf \normalsize INTRODUCTION}

The viscoplastic deformation of crystalline materials, such as ice
single crystals, involves the motion of a large number of
dislocations. Although the dynamics of an individual dislocation is a
fairly well characterized phenomenon~\cite{Hirth92,Nabarro87}, the
collective behavior of a large number of these defects appears to be
an amazingly rich but poorly understood problem. The interaction
between a pair of dislocations can be attractive or repulsive,
depending on the orientation of their respective Burgers vectors; it
grows logarithmically with the interline distance and allows mutual
annihilation of defects. Dislocations may be incorporated into a
crystal in the growth process, affecting the topology of the whole
lattice.  Moreover, under deformation conditions, dislocations can
penetrate into the material from the sample surfaces, or be generated
by various mechanism, as for example, in what is usually called a
Frank-Read source, which is activated after the pinning of a
dislocation loop.

When a material is deformed under constant load (creep experiment),
and dislocation motion is the dominant mechanism for viscoplastic
deformation (other possible source is for instance crack nucleation
and propagation), a constant strain-rate regime usually follows after
the initial transient stage. Orowan's relation $\dot \gamma = \rho_m b
v$ is known to prevail under such conditions, where $\gamma$ is the
strain of the sample, $\rho_m$ is the density of mobile dislocations,
$b$ is the Burgers' vector, and $v$ is the mean velocity of the
dislocations. Obviously, this is a mean-field relation which neglects
temporal and spatial fluctuations of both the density and the velocity
fields. As a result of their interactions, however, dislocations tend
to move cooperatively giving rise to a rather complex and
heterogeneous slip process. Dislocations move in groups to form slip
bands. Moving dislocations can pile-up against stable dislocation
configurations such as walls or boundaries, which may eventually break
apart. In this process, fluctuations in the dislocation density and
velocity may be comparable to or greater than the mean values, and
consequently, of great importance.

The complex character of the collective dislocation dynamics reveals
itself in experiments of acoustic emission (AE)~\cite{Weiss97}. Sudden
local changes of inelastic strain generate AE waves. Weiss and
Grasso~\cite{Weiss97} soon realised that ice single crystals are
particularly well suited for the study of dislocation dynamics. The
perfect transparency of this material, easily (by eye) allows to rule
out the possibility of cracks being present in the material which
will, otherwise, interfere with the dislocations motion. Given the
amplitude threshold and the frequency range accesible to the
experimental apparatus, the AE signals detected seem to correspond to
the synchronous motion of several dislocations, likely to occur for
example during the breakaway of a pile of these defects, or the
activation of a multiplication source. The AE experiments, however,
have only access to information resulting from the interplay of
various magnitudes. Thus, the physical interpretation of the generated
AE waves remains a major difficulty, and consequently, constitutes the
main motivation of our work.

The proportionality between global AE activity and global strain-rate
(Orowan's relation) has been tested experimentally. More locally, the
AE intensity is thought to depend on the number of dislocations
involved in a plastic instability, their length, and velocity. Various
measurements of the acoustic activity recorded during a
stress-constant step show that the AE signal takes place in the form
of bursts spanning a wide range of amplitude values. In particular,
the AE amplitude is distributed according to a power law. This
behavior is a consequence of the collective motion of dislocations
which spontaneously gives rise to an avalanche-like dynamics, typical
of slowly driven dissipative systems. The power law distribution
provides evidence of scale-free cooperative behavior, whose origin
could be ascribed to nonequilibrium continuous phase
transitions~\cite{transitions} or self-organized
criticality~\cite{soc}.

\section*{\bf \normalsize DESCRIPTION OF THE MODEL}

To characterize the plastic deformation of a material from the
perspective of nonequilibrium statistical mechanics, we propose a
simplified dynamic model to simulate the system in rather general
conditions.

Ice single crystals deform essentially by slip on the basal plane
$(0001)$ (that we call the $xy$ plane), i.e. the motion of dislocation
lines or loops takes place by gliding on the $xy$ planes. The possible
Burgers vectors ${\bf b}$ are the three lattice vectors of an
hexagonal lattice. For the sake of simplicity, we study a
two-dimensional model representing a cross section of the crystal
which is perpendicular to the basal planes and parallel to one of the
lattice vectors, that is, for example, the $xz$ plane. In this way,
the dislocations constrained to move in this plane have all Burgers
vectors parallel to the chosen lattice vector ${\bf b}=(b,0,0)$ and
move along fixed lines parallel to the $x$ axis.  We also consider
that all $N$ dislocations are of edge type, and that, on average over
many realizations, the number of dislocations with positive and
negative Burgers vectors is the same.

Several simplified models containing similar basic ingredients have
been proposed in the literature in the last few
years~\cite{Kubin87,Amodeo90,Groma93,Fournet96,Miguel99}. A basic
feature common to most models, is that dislocations interact with each
other through the long-range elastic stress field they produce in the
host material. An edge dislocation with Burgers vector $b$ located at
the origin gives rise to a shear stress $\sigma^{s}$ at a point
$(x,z)$ of the form

\begin{equation}\label{eq:1}
\sigma^{s}=bD\frac{x(x^2-z^2)}{(x^2+z^2)^2},
\end{equation} 

\noindent where $D=\mu/2\pi(1-\sigma)$ is a coefficient involving the
shear modulus $\mu$ and the Poisson's ration $\sigma$ for the
material.  In our model, we further assume that the dislocation
velocities are linearly proportional to the local stress. Experimental
evidence supports such a relationship for low stress conditions, which
is indeed the case in our model. Accordingly, the velocity of the
$n$th dislocation, if an external shear stress $\sigma^{e}$ is also
applied, is given by

\begin{equation}\label{eq:2}
v_n=b_n(\sum_{m\neq n} \sigma^{s}_{nm} - \sigma^{e}).
\end{equation} 

As the number of dislocations in any real crystals exceeds by far the
number of defects we can handle in a computer, one usually introduces
periodic boundary conditions (PBC) to effectively extend the size of
our system. Due to the long range character of the force (\ref{eq:1}),
we have exactly evaluated the Ewald sums of this expression to account
for the interaction of a dislocation with all the infinite periodic
replicas of all the other dislocations in a finite box of dimensions
$L\times L$. Contrary to what is stated in some
references~\cite{Groma93}, we do not find any spureous results coming
from the implementation of PBC in this fashion. Instead, we obtain
artificial results when using the ``nearest image'' approximation and
the truncation that this approximation implies.

When the distance between two dislocations is of the order of a few
Burgers vectors, the high stress and strain conditions close to the
dislocation core invalidate the results obtained from a linear
elasticity theory (i.e. Eq~.(\ref{eq:1})). In these instances,
phenomenological nonlinear reactions describe more accurately the real
behavior of dislocations in a crystal. In particular in our model, we
account for the {\em annihilation} of dislocations with opposite
Burgers vectors when the distance between them is shorter than
$2b$. Thus the core of one dislocation in our model has a radius
of size $b$. 

Another important feature of any computer model is the implementation
of a mechanism for the {\em multiplication} of dislocations in the
sample. It is widely believed that the Frank-Read
mechanism~\cite{Hirth92} is the most relevant for a gliding process of
dislocations under creep deformation. Indeed Frank-Read sources (FRS)
have been observed in ice. In a FRS multiplication occurs by pinning
of a dislocation segment on the basal planes due, for example, to a
defect in the crystal, or to dislocation dipoles, piles, and
walls. Under an applied stress, the pinned segment bows out by glide
and, if the local stress concentration is less than a critical value,
a metastable configuration is attained where the line tension balances
the stress. Beyond this threshold value, the dislocation segment wraps
around itself, creating a new dislocation loop and restoring the
original configuration. Thus a sequence of loops forms continuously
from the source until the local shear stress drops below the
activation value.  In our model, we simulate this mechanism
phenomenologically: a dislocation pair is generated (i) when the
fraction of immobile dislocations is high (pinning is then more
likely) and (ii) the local stress is large compared to a threshold
value. Rather than fixing particular values of the parameters, we use
a probabilistic procedure, keeping in mind that the details of the
rule should not change the collective properties of the system.

Annihilation and multiplication processes imply that the number of
dislocations $N$ in our system is not fixed in the course of
time. Starting from a random configuration of dislocations, we let
them relax until they find a stable arrangement (which could be an
equilibrium configuration, or a long-lived metastable state). We solve
numerically the $N$ equations of motion using an {\em adaptive step
size fifth-order Runge-Kutta method}. So far, we have considered three
different box sizes $L=100b$, $L=200b$, and $L=300b$, with an initial
number of dislocations $N_0=400$, $N_0=800$, and $N_0=1500$,
respectively. After the system has reached a stable arrangement, the
volume fraction of dislocations $\phi=N \pi b^2/(Lb)^2$ ranges between
$1-10$\%. Once in these conditions, we apply a small external shear
stress and keep track of the various quantities describing the
dynamics of the dislocations.

\begin{figure}
\centerline{\epsfig{file=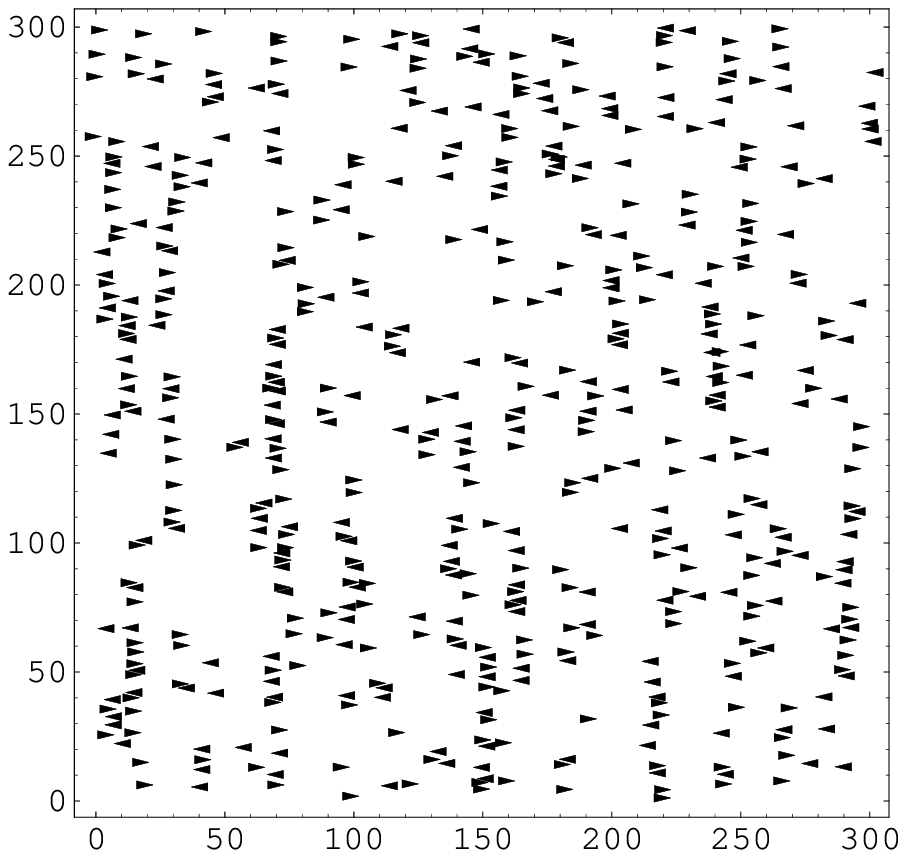, width=8truecm}
\epsfig{file=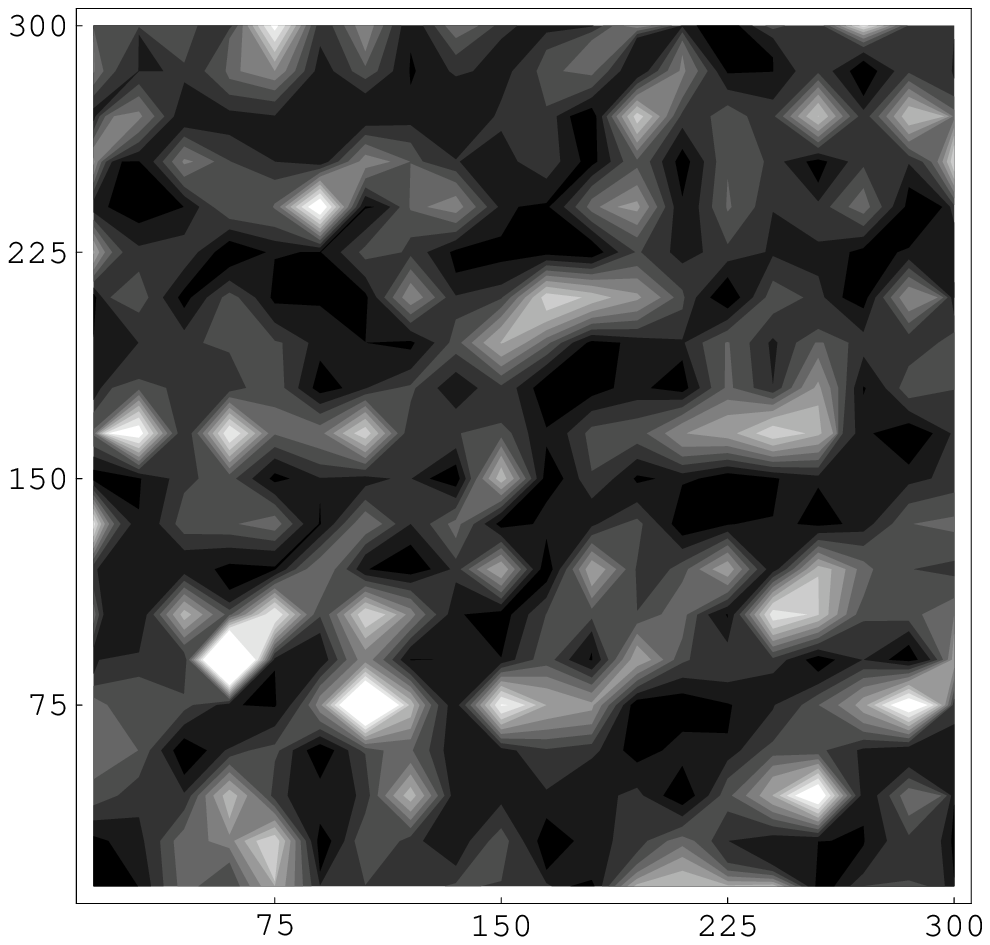, width=8truecm}}
%\medskip
\vspace{-0.25truecm}
\caption{\small a) A stable arrangement of dislocations in a box of
size $L=300b$. Various structures like dislocation dipoles, piles, and
dislocation walls are present. b) The corresponding elastic stress map
(absolute value) in a graylevel scale. Dark color indicates a
low-stress region and lighter colors areas of higher stress.
\label{patterns}}
\vspace{-0.25truecm}
\end{figure} 

\section*{\bf \normalsize RESULTS AND DISCUSSION}

Persistent slip bands or cell structures appearing on various length
scales, are just a few examples of dislocation patterns.  With our
approach we are able to recover some of these structures. This is a
subject which has raised a considerable amount of activity recently
(see for example Refs.~\cite{Hahner98,Bako99} and references therein).
Fig.~\ref{patterns}a) represents a common configuration of
dislocations in stable conditions.  There, we can observe the
formation of several dipoles, piles, as well as dislocation walls
delimiting various slip bands.  The corresponding elastic stress map
is depicted in Fig.~\ref{patterns}b). We plot the absolute value of
the shear stress in a graylevel scale. The dark areas represent
low-stress regions, and are consequently predominant in a stable
configuration. The light portions mark a few regions of higher stress
concentration given this particular arrangement.

\begin{figure}
\centerline{\epsfig{file=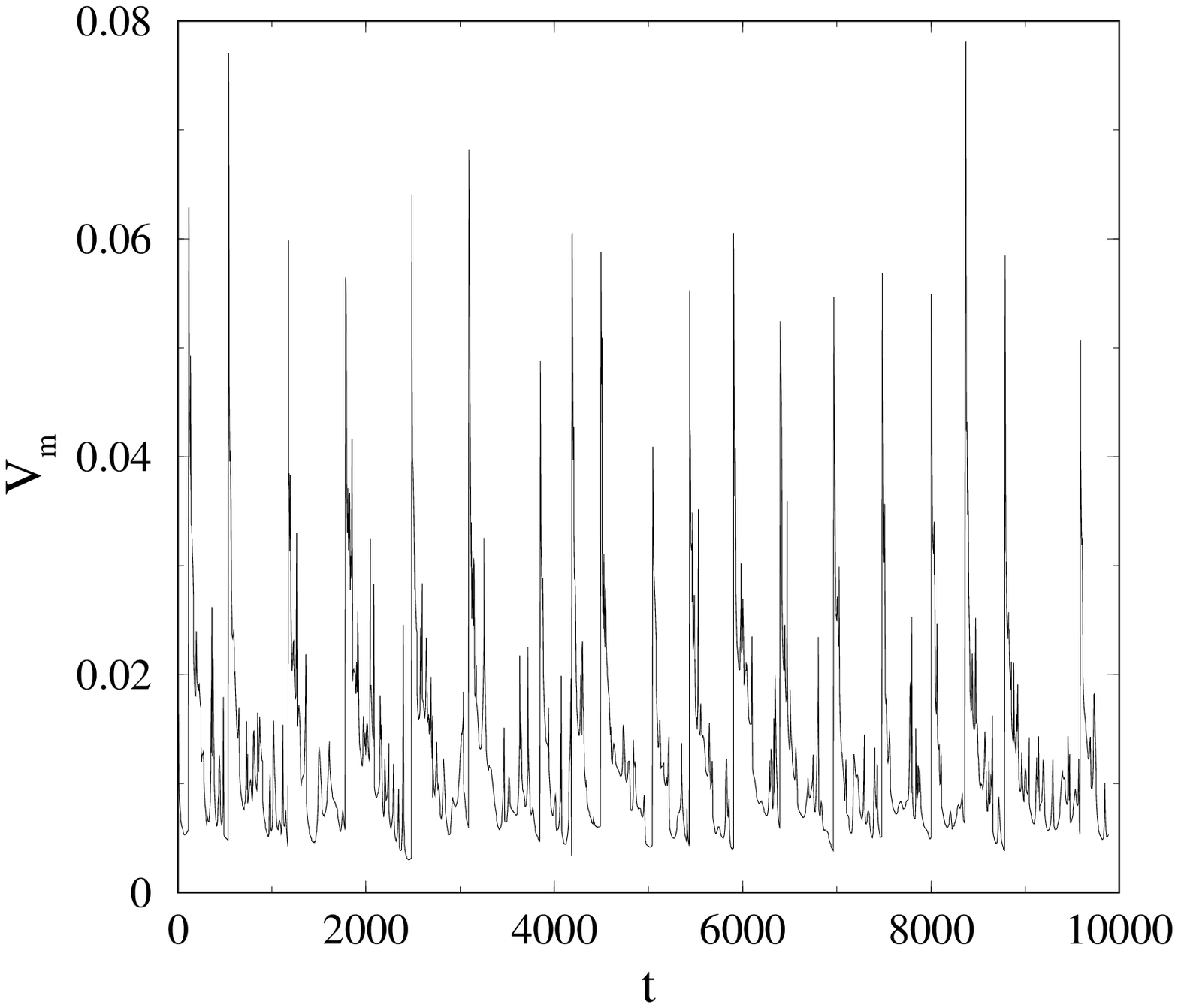,width=8truecm}
            \epsfig{file=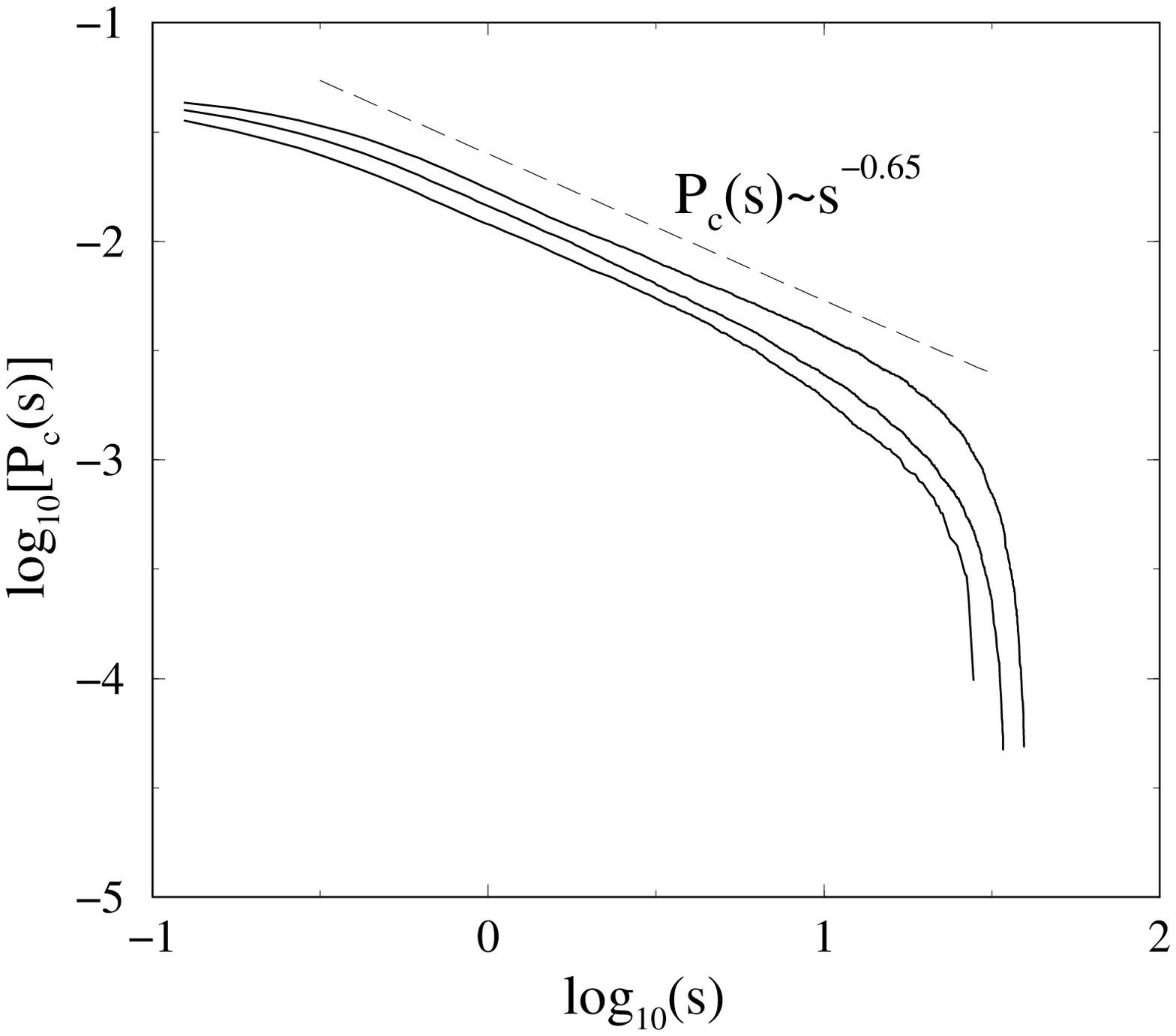, width=8truecm}} 
\vspace{-0.5truecm}
\caption{\small a) Mean velocity as a function of time in our model of
collective dislocation dynamics. b) The cumulative {\em avalanche}
size distribution for three system sizes $L=100,200,300$ represented
in a double logarithmic scale.  \label{velo} }
\vspace{-0.25truecm}
\end{figure}

Second, we have been able to keep track of several quantities which
play a key role in promoting scale invariant behavior, like the
root-mean-square velocity, the local stress, the average number of
dislocations, etc. In Figure~\ref{velo}a), we represent, for example,
the root-mean-square velocity $V_m$ of a single run of the creep
process. A curious feature is that after a burst, the relaxation of
the velocity $V_m$ to a still configuration is power-law like,
i.e. slow and without any characteristic time.

We have defined the size $s$ of an acoustic {\em avalanche} as the sum
of the root-mean-square velocity of all the moving dislocations, that
is $s=\sum |v_i|$. The cumulative distribution $P_c(s)$ of the
avalanches obtained after averaging over several realizations is
depicted in Figure~\ref{velo}b) for the three system sizes studied
$L=100,200,300$.  We recover a very clear power law distribution
($P_c(s)\sim s^{-\tau}$) extending over close to two decades. The
exponent $\tau\simeq 0.65$ is in reasonable agreement with
experimental data~\cite{Weiss97}.  The distribution cut-off for large
values of $s$ is due to the finite size of the sample. (As one would
expect, the cut-off is scaling accordingly to the sytem size.) This
clearly points out the presence of a very large (or infinite)
characteristic size for the acoustic events. It is worth remarking
that large stresses introduce a characteristic scale in the process. A
more detailed study as a function of the applied stress is in
progress~\cite{Miguel99}.

The numerical investigation of the present model provides striking
evidences for the collective critical behavior of dislocation motion
under external stress. Relevant magnitudes characterizing the dynamics
of dislocations show power law behavior signalling the absence of any
characteristic length or time in the process. The response to an
infinitesimal perturbation (slow injection of new dislocations)
exhibits singular behavior in the guise of avalanches distributed over
many length scales.  Avalanche dynamics is the rule rather the
exception in slowly driven disordered systems.  Examples can be found
in fracturing of wood and concrete~\cite{frac}, Barkhausen
effect~\cite{durin94}, and flux lines in high-$T_c$
superconductors~\cite{nori95}.  Under the external drive (the stress
in the present case), the system jumps between metastable or pinned
configurations in which the dynamics is virtually frozen. In the limit
of a very slow driving (that is equivalent to a fine tuning close to
the depinning critical point), the disordered energy landscape is
explored quasistatically and the response function exhibits critical
properties~\cite{ron99}.  Tipically, a basic ingredient for this
behavior is the presence of quenched disorder acting as the source of
pinning in the system.  Noticeably, the system under study does not
contain any external source of disorder. Pinned states are due to the
various dislocation patterns that spontaneously develop in the
system. Structures such as dipoles, piles, and dislocation walls, play
the role of self-generated pinning centers that create the pinning
force landscape.  The new scenario poses many new and interesting
questions for a definitive identification and understanding of the
critical nature of dislocation dynamics.

\section*{\bf \normalsize ACKNOWLEDGMENTS}

We gratefully acknowledge R. Pastor-Satorras advice on the algorithm
implementation. We also thank J.R. Grasso and J.Weiss for fruitful
discussions regarding their experiment, and for providing us with
the experimental data prior to publication.

\end{document}